\documentclass[nobibnotes,prb,showpacs,twocolumn]{revtex4}
\usepackage{amsmath,amssymb,bm,graphicx}

\newcommand{\bto}{BaTiO$_3$}
\newcommand{\magnetite}{Fe$_3$O$_4$}

\begin{document}

\title{Magnetoelastic coupling in \bto-based multiferroic structures}

\author{C. A. F. Vaz}
\email[Email: ]{carlos.vaz@yale.edu} \affiliation{Department of
Applied Physics, Yale University, New Haven, Connecticut 06520}%
\affiliation{Center for Research on Interface Structures and
Phenomena (CRISP), Yale University, New Haven, Connecticut 06520}%

\date{\today}

\begin{abstract}
Analytical expressions for the magnetoelastic anisotropy constants
of cubic magnetic systems are derived for rectangular and oblique
distortions originating from epitaxial growth on substrates with
lower crystal symmetry. In particular, the temperature variation of
the magnetic properties of magnetic films grown on barium titanate
(BaTiO$_3$) substrates are explained in terms of strain-induced
magnetic anisotropies caused by the temperature dependent phase
transitions of BaTiO$_3$. Our results quantify the experimental
observations in ferromagnet/\bto-based structures, which have been
proposed as templates for magnetoelectric composite
heterostructures.
\end{abstract}



\maketitle

\section{Introduction}

Complex oxides are characterized by a strong coupling between
charge, spin and lattice distortions, giving rise to a rich variety
of electronic and magnetic behavior which is at the origin of the
multifunctional properties exhibited by this class of materials. In
particular, much effort is being made towards the control of these
properties by means of electric and magnetic fields. While intrinsic
multiferroic materials already exhibit a coupling between magnetic
and electric order
parameters,\cite{Skinner70,SC82,Schmid94,Fiebig05} the magnitude of
the coupling in these compounds is weak, and a new class of
composite materials is being developed that exploit the different
properties of the constituent compounds to tailor systems with
larger susceptibilities and more varied response functions than
those of single component systems.

In this context, the cross coupling in ferroelectric/magnetic
heterostructures and composite materials is being intensively
studied. The coupling mechanism can be classified as strain-driven
or charge driven: in the first case, the piezoelectric effect of
ferroelectrics is used to modify the magnetic properties of ferro-
or ferrimagnets via magnetoelastic
coupling;\cite{MW02,ZWL+04,Fiebig05,ZZM+05,TDB+07} in the second
case, the coupling originates from modifications in the electronic
properties of the magnetic system driven by charge modulation
induced by the polarization state of the ferroelectric.\cite{MHV+08}

A particular case of strain-mediated magnetoelectric coupling occurs
in epitaxial heterostructures composed of thin magnetic films grown
on \bto\ single crystals, including
La$_{1-x}$Sr$_{x}$MnO$_3$,\cite{LNE+00,DFBS03,EWP+07}
SrRuO3,\cite{LNE+00} CoFe$_2$O$_4$,\cite{CS06} Fe,\cite{SPD+07} and
\magnetite,\cite{ZBPM06,TQL+08,VHPA08} where the changes in strain
occur as a function of temperature via the phase transitions of
\bto\ or via applied electric fields. However, to date, the
magnetoelastic coupling has not been addressed explicitly in detail.
Here, we present analytical expressions for the magnetoelastic
contributions to the magnetic anisotropy induced by elastic coupling
between the different phases of the \bto\ substrate to the epitaxial
magnetic film. This coupling explains the magnetic behavior
exhibited by these systems as a function of temperature in terms of
the changes in the magnetoelastic anisotropies.

\bto\ has a simple perovskite structure at high temperatures and
undergoes several crystal phase transitions as a function of
temperature.\cite{JS62,AAA+02} It is cubic in the temperature range
from $\sim$1733 K to $\sim$393 K ($a = 3.996$ at 393 K), tetragonal
at temperatures down to $\sim$278 K ($a=3.9920$ \AA\ and $c=4.0361$
\AA\ at 293 K), orthorhombic down to $\sim$183 K
($a_\mathrm{o}=5.669$ \AA, $b_\mathrm{o}= 5.682$ \AA, and
$c_\mathrm{o} = 3.990$ \AA\ at $T=263$ K), and rhombohedral at lower
temperatures ($a = 4.001$ \AA, $\gamma=89.85^\mathrm{o}$ at 105 K).
For temperatures below $\sim$393 K, \bto\ is ferroelectric. The
non-cubic lattices correspond to slight distortions of the cubic
phase, with the direction of the distortion linked to that of the
electric polarization, {\bf P}:\cite{JS62,AAA+02} in the tetragonal
phase the electric polarization points along the $c$ axis; in the
orthorhombic phase, the spontaneous polarization points along the
$\langle 110\rangle$ directions of the cubic cell, leading to a
slight elongation of the unit cell along that direction. Although
the symmetry group is orthorhombic, with lattice parameters
$a_\mathrm{o}$, $b_\mathrm{o}$, and $c_\mathrm{o}$ when {\bf P}
points along [110], it is often convenient to consider the
monoclinic cell with
$a_\mathrm{m}=b_\mathrm{m}=(a_\mathrm{o}^2+b_\mathrm{o}^2)^{1/2}/2$
and $\gamma= 2\arctan(a_\mathrm{o}/b_\mathrm{o})$, giving $a = b =
4.013$ \AA, $c = 3.99$ \AA, $\gamma=89.869^\mathrm{o}$ at $T=263$ K.
For the rhombohedral phase, the polarization points along the
$\langle 111\rangle$ directions of the cubic cell, corresponding to
an elongation of the cubic cell along {\bf P}.

\section{Magneto-elastic energy expressions}

In the following, we are interested in determining the effect of the
in-plane strain on the anisotropy constants of ferromagnetic (or
ferrimagnetic) films due to magnetoelastic coupling. We assume that
the strain propagates uniformly throughout the film. When relaxation
occurs, as is often the case in real systems, the expressions below
correspond to upper bounds to the magnetolastic anisotropy induced
in the magnetic system. If the residual strain is known, the
effective anisotropies can be estimated.

To calculate the magnetoelastic anisotropy contributions arising
from strain, we minimize the sum of the elastic ($F_\mathrm{elas}$)
and magnetoelastic ($F_\mathrm{me}$) energies for a given strain
state of the system. We consider the case of a cubic magnetic
system:\cite{Love20,Carr66,Chikazumi97}
\begin{eqnarray}
F_\mathrm{me} & = & b_1\sum_i\alpha_i^2\epsilon_{ii} +
b_2\sum_{i>j}\alpha_i\alpha_j\epsilon_{ij}\\
F_\mathrm{elas} & = & \frac{c_{11}}{2}\sum_i\epsilon_{ii}^2 +
c_{12}\sum_{i>j}\epsilon_{ii}\epsilon_{jj} +
\frac{c_{44}}{2}\sum_{i>j}\epsilon_{ij}^2,
\end{eqnarray}
where $\alpha_i$ are the direction cosines of the magnetization,
$\epsilon_{ij}$ are the components of the strain tensor, $b_i$ are
the magnetoelastic coupling coefficients, $c_{ij}$ are the elastic
constants and $i,j$ are the cartesian coordinate indexes. The
magnetoelastic constants are related to the magnetostriction
coefficients by:\cite{Chikazumi97}
\begin{eqnarray}
b_1& = & -3(c_{11}-c_{12})\lambda_{100}/2\\
b_2& = & -3c_{44}\lambda_{111}
\end{eqnarray}

\begin{figure}[t!b]
\centering
\includegraphics[width=.6\columnwidth]{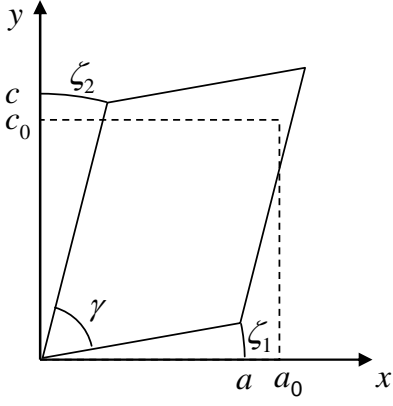}
\caption{Schematic of a general (oblique) distortion of a
rectangular unit cell. The strain components correspond to the
normalized difference between the distorted and the original
equilibrium unit cell.} \label{fig:strain}
\end{figure}

For a general in-plane distortion in a cubic lattice, it is
straightforward to calculate the in-plane strain tensor components,
since by definition the latter correspond to the relative deviations
from the equilibrium lattice constant; with the aid of
Fig.~\ref{fig:strain}, we find:
\begin{eqnarray}
\epsilon_{xx} & = & (a\cos\zeta_1 - a_0)/a_0\\
\epsilon_{yy} & = & (c\cos\zeta_2 - c_0)/c_0\\
\epsilon_{xy} & = & [(a/a_0)\sin\zeta_1 + (c/c_0)\sin\zeta_2]/2,
\end{eqnarray}
where $a_0$, $c_0$ are the equilibrium in-plane lattice parameters
of the magnetic film and $a$, $c$, $\gamma$ and $\zeta$ are the
parameters characterizing the planar lattice distortion. Here, we
discuss the case of a (001) ferromagnetic film, where $a_0=c_0$, but
the calculations can be easily modified for other cases. The out of
plane strain components are free to settle to the values that
minimize the total energy (zero out of plane stress). The
magnetoelastic anisotropy is then calculated by finding the minimum
of $F_\mathrm{m} = F_\mathrm{me} + F_\mathrm{elas}$ with respect to
$\epsilon_{xz}$, $\epsilon_{yz}$ and $\epsilon_{zz}$:
\begin{eqnarray}
\label{eq:ezz} %
\epsilon_{zz} & = &
-\frac{c_{12}}{c_{11}}(\epsilon_{xx}+\epsilon_{yy})- \frac{b_1}{c_{11}}\alpha_z^2\\
\label{eq:exz} %
\epsilon_{xz} & = & -\frac{b_2}{c_{44}}\alpha_x\alpha_z\\
\label{eq:eyz} %
\epsilon_{yz} & = &  -\frac{b_2}{c_{44}}\alpha_y\alpha_z;
\end{eqnarray}
inserting (\ref{eq:ezz}-\ref{eq:eyz}) into the expression for
$F_\mathrm{m}$ gives, retaining only the terms that depend on the
magnetization:
\begin{eqnarray}
\nonumber
 F_\mathrm{m} & = & b_1\left[\epsilon_{xx}\alpha_x^2 + \epsilon_{yy}\alpha_y^2 -
 \frac{c_{12}}{c_{11}}(\epsilon_{xx}+\epsilon_{yy})\alpha_z^2\right] + \\
 & & + b_2\epsilon_{xy}\alpha_x\alpha_y + o(b^2/c)
\end{eqnarray}
where the last term is strain independent, corresponding to a sum of
terms of the order of $b_1^2/c_{11}$ and $b_2^2/c_{44}$ with
$\alpha_z^2$ and $\alpha_z^4$ dependencies. These are much smaller
than the first terms and will be ignored in the following
discussion. They arise from a shear strain exerted by the
magnetization on the lattice (which favors specific directions that
depend on the relative sign of the magnetoelastic coupling
coefficients) and are usually neglected to first order of
approximation. We note that the magnetoelastic terms are uniaxial
and lower in symmetry than the magnetocrystalline anisotropy of the
unstrained cubic magnetic system.

A more convenient expression is obtained by using spherical
coordinates, where the polar axis is set along the out of plane
direction:
\begin{eqnarray}
\nonumber
 F_\mathrm{m} & =& \frac{1}{2}[b_1(\epsilon_{xx}-\epsilon_{yy})\cos 2\phi + b_2\epsilon_{xy}\sin 2\phi ]\sin^2\theta -\\
 & & - \frac{b_1}{2}(1+2c_{12}/c_{11})(\epsilon_{xx}+\epsilon_{yy})\cos^2\theta
\end{eqnarray}
where the first two terms correspond to in-plane magnetoelastic
anisotropies, and the last term is the perpendicular magnetoelastic
anisotropy contribution. An in-plane uniaxial anisotropy arises for
uniaxial in-plane strains ($\epsilon_{xx}\ne \epsilon_{yy}$) or for
shear strains, although the anisotropy direction is distinct for
these two cases: in the first case, it is oriented along the main
crystal axes, while for shear strains, the anisotropy is directed
along the in-plane diagonal directions. A perpendicular anisotropy
arises when $\epsilon_{zz} \ne 0$. The relative magnitude of
$b_1\epsilon_{zz}$ with respect to the magnetostatic energy term
(shape anisotropy, which favors in-plane magnetization), which is on
the order of $2\pi M_s^2$, where $M_s$ is the saturation
magnetization, determines whether a state of in-plane or
perpendicular magnetization is favored. For the particular case of a
biaxial strain ($\epsilon_{xx} = \epsilon_{yy}$, $\epsilon_{xy}=0$),
the magnetoelastic anisotropy energy reduces to a perpendicular
magnetic anisotropy term only, with
$K_\mathrm{p}=-b_1(1+2c_{12}/c_{11})\epsilon_{xx}$, as
expected.\cite{VBL08}

We consider next the effect of square, rectangular, oblique and
rhomboidal distortions on the magnetoelastic anisotropy. These
planar domains arise in the cubic, tetragonal, orthorhombic and
rhombohedral phases of BaTiO$_3$, and are responsible for the large
magnetic anisotropy modulations observed in
ferromagnetic/BaTiO$_3$(001) heterostructures. We consider the
magnetoelastic anisotropy constants, $K_\mathrm{t}$, $K_\mathrm{o}$,
$K_\mathrm{p}$, expressed as:
\begin{equation}
F_\mathrm{m} = [K_\mathrm{t}\cos 2\phi + K_\mathrm{o}\sin 2\phi
]\sin^2\theta + K_\mathrm{p}\cos^2\theta.
\end{equation}
The direction of easy magnetization depends on the particular $K$
values; positive $K_\mathrm{p}$ favors in-plane magnetization. The
magnetoelastic anisotropy for the cubic phase of \bto\
($\zeta_1=\zeta_2=0$, $a=c$) leads to zero in-plane anisotropy. For
the tetragonal phase ($\zeta_1=\zeta_2=0$), two types of domains are
possible, $c$-domains (square) and $a$-domains (rectangular, which
may co-exist in two orthogonal directions). A multidomain state can
arise when cooling the \bto\ from the cubic to the tetragonal phase
after film growth at elevated temperatures, for example. For this
phase, no shear strains are present and $K_\mathrm{o} =0$. The
orthorhombic phase may give rise to (two) rectangular
($\zeta_1=\zeta_2 =0$, $a\ne c$) and (two) oblique
($\zeta_1=\zeta_2$, $a = c$) domains. Finally, the rhombohedral
phase ($\zeta_1=\zeta_2$, $a=c$) is characterized by (two) oblique
domains. It also follows that for the oblique domains of the
orthorhombic and rhombohedral phases of \bto, $K_\mathrm{t}=0$.

\section{Application to epitaxial magnetic films}

In general, it may be difficult to control the ferroelectric state
of \bto\ across the different phase transitions. For example, the
growth of oxide magnetic films is often carried out at elevated
temperatures, where \bto\ is cubic and, as the temperature is
lowered to room temperature, the \bto\ tends to break into a
ferroelectric multidomain state, giving rise to $a$ and $c$ domains
in a \bto(001) single crystal. Similar processes operate at the
other phase transitions, leading to the presence of different types
of planar domains at the interface with the magnetic film. In such
cases, an effective magnetoelastic anisotropy corresponds to an
average over the relative number and size of domains present; in the
case where an equal number of the two possible oblique domains are
present in the rhombohedral phase, no effective in-plane anisotropy
would result. However, experimental estimates of the relative
population of the different domains is difficult. In the following
we will consider single domain states; for similar, but orthogonal,
domains, the in-plane anisotropy constants will be identical, but
with opposite signs, while the perpendicular anisotropies are
unaffected. Another difficulty relates to the fact that for systems
with large lattice mismatch, such as Fe$_3$O$_4$ and CoFe$_2$O$_4$
grown on \bto, the strain is strongly reduced for films thicker than
the coherent critical thickness through the onset of misfit
dislocations;\cite{Matthews75,HB84} in the case of
Fe$_3$O$_4$/\bto(100), the experimental evidence points to full
relaxation at the temperature of growth.\cite{VHPA08} Here, we do
not take explicitly into account the effect of dislocations to the
magnetoelastic energy, but instead account for strain relaxation by
adjusting the lattice parameter of the magnetic film so that the
correct strain value is obtained at the measurement temperature, as
will be discussed in more detail below. We consider three magnetic
systems that have been studied experimentally and which are of
particular interest: CoFe$_2$O$_4$,\cite{CS06} Fe$_3$O$_4$,
\cite{VHPA08,TQL+08} and Fe films\cite{SPD+07} grown on \bto. In
Table~\ref{tab:constants}, we list the relevant materials parameters
for these magnetic systems.

\begin{table*}[hbt]
\caption{Room temperature material parameters for Fe, Fe$_3$O$_4$
and CoFe$_2$O$_4$. The elastic constants $c_{ij}$ are given in
erg/cm$^3$, the magnetization $M_s$ in emu/cm$^3$, and the cubic
magnetocrystalline anisotropy $K_1$ in erg/cm$^3$ ($a$ is the
lattice constant).}\label{tab:constants}
\begin{ruledtabular}
\begin{tabular}{lllllllll}
Material & $a$ (\AA) & $10^{-12}c_{11}$ &  $10^{-12}c_{12}$ &  $10^{-12}c_{44}$ & $10^{6}\lambda_{100}$ & $10^{6}\lambda_{111}$ & $M_s$ & $10^{-6}K_1$\\ \hline %
Fe\footnotemark[7] & 2.866 & 2.322 & 1.356 & 1.170 & 24.1 & -22.7 & 1717 & 0.48 \\ %
Fe$_3$O$_4$\footnotemark[1] & 8.397 & 2.73 & 1.06 & 0.97 & -19.5 & 77.6 & 471 & -0.110\\ %
CoFe$_2$O$_4$ & 8.381\footnotemark[2] & 2.57\footnotemark[3] & 1.500\footnotemark[3] & 0.853\footnotemark[3] & -590\footnotemark[4] & 120\footnotemark[4] & 425\footnotemark[5] & 1.8\footnotemark[6] \\ %
\end{tabular}
\end{ruledtabular}
\footnotetext[1]{From Ref.~\onlinecite{Lefever70}.} %
\footnotetext[2]{From Ref.~\onlinecite{PT80}.} %
\footnotetext[3]{From Ref.~\onlinecite{LFLN91}.} %
\footnotetext[4]{From Ref.~\onlinecite{Folen70}, for single crystalline Co$_{0.8}$Fe$_{2.2}$O$_4$.} %
\footnotetext[5]{From Ref.~\onlinecite{Folen70}.}
\footnotetext[6]{From Ref.~\onlinecite{Folen70}, for sample with composition Co$_{1.1}$Fe$_{1.9}$O$_4$.} %
\footnotetext[7]{From Ref.~\onlinecite{Stearns86}.} %
\end{table*}

\subsection{Fe$_3$O$_4$/\bto(100) epitaxial films}

\begin{figure}[t!]
\includegraphics[width=\columnwidth]{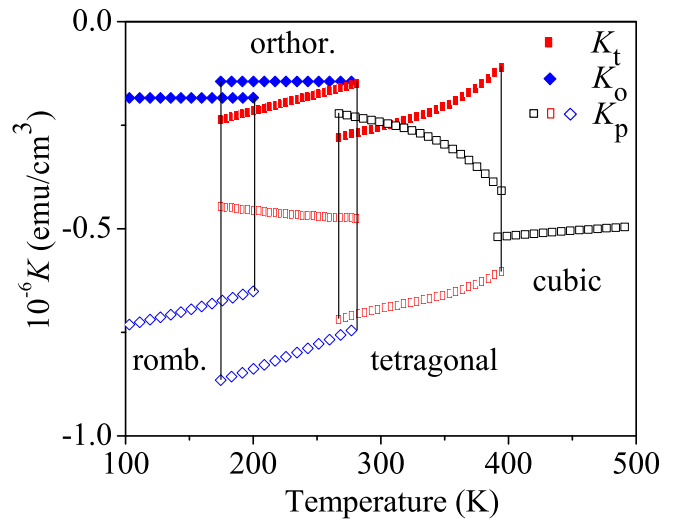}
\caption{Variation of the magnetoelastic anisotropy constants of
partially strained Fe$_3$O$_4$/\bto(001). Square symbols correspond
to anisotropy in planar square domains, diamonds to oblique domains,
and rectangles to rectangular domains; full symbols are for in-plane
anisotropies, empty symbols for the perpendicular anisotropy
constants.} \label{fig:Fe3O4}
\end{figure}

We consider first the case of Fe$_3$O$_4$/\bto(100) epitaxial films,
where experimentally it is found that the Fe$_3$O$_4$ has a residual
strain at room temperature with a sign that is opposite that
expected from the lattice mismatch with \bto; this has been
explained as due to strain relaxation at the temperature of growth,
followed by an in-plane dilation due to the abrupt increase in the
cell parameters of \bto\ at the cubic to tetragonal phase
transition.\cite{VHPA08} Therefore, the magnetoelastic anisotropy
contribution to the magnetic energy is very different from that
expected for a fully strained Fe$_3$O$_4$ film: while the values for
the magnetostriction coefficients of Fe$_3$O$_4$ favor a state of
in-plane magnetization, experimentally a non-zero perpendicular
remanence and magnetic anisotropy are observed.\cite{TQL+08,VHPA08}
In Fig.~\ref{fig:Fe3O4} we present our estimate for the variation of
the magnetoelastic anisotropy constants as a function of
temperature, where we took into account the temperature variation of
the lattice constants of \bto\ and
Fe$_3$O$_4$.\cite{Shebanov81,YI79} The strain relaxation was taken
into account, in a first approximation, by shifting the lattice
constant of magnetite such that the out of plane strain at room
temperature coincides with the experimental value of --0.62\% for
tetragonal domains.\cite{VHPA08} This assumes that the misfit
dislocation density remains constant throughout the temperature
range and across the phase transitions of \bto. One indication that
this approach is a simplification over the real processes occurring
in the magnetite film is the fact that, as observed in
Fig.~\ref{fig:Fe3O4}, the strain in the cubic phase does not drop to
zero as the temperature approaches the growth temperature,
indicating that a strong temperature dependence of the lattice
constant, or misfit dislocation density, may occur in the high
temperature range. The temperature variation of the magnetic
anisotropies is instructive, however, and helps explain the
variations in the magnetic properties observed experimentally.
Focusing on the out of plane anisotropy constant $K_\mathrm{p}$
first, we see that negative out of plane strains favor perpendicular
anisotropy, which are large for the rectangular domains of the
tetragonal phase and for the oblique domains of the orthorhombic and
rhombohedral phases; they are smaller by a factor of about 2 for the
square domains of the tetragonal phase and the rectangular domains
of the orthorhombic phase. Depending on the particular domain path
followed by the \bto\ substrate, it is clear that large changes in
the magnetic anisotropies may follow. For example, the experimental
results of Vaz et at.\cite{VHPA08} can be explained if, starting
from a rectangular domain at room temperature, the \bto\ surface
reverts to a rectangular domain in the orthorhombic phase,
corresponding to a drop in the perpendicular magnetic anisotropy,
followed by an increase in $K_\mathrm{p}$ in the rhombohedral phase.
The in-plane anisotropies are also significant, but they tend to be
smaller than the perpendicular anisotropies. The presence of
multidomains in \bto\ also tends to average out the in-plane
effective anisotropy.

It must be emphasized that a negative $K_\mathrm{p}$ is not
sufficient to yield perpendicularly magnetized films, since the
in-plane magnetostatic (shape) anisotropy must also be overcome, and
the latter is on the order of $2\pi M_s^2 = 1.39 \times 10^6$
erg/cm$^3$ for magnetite, assuming bulk magnetization (thin
magnetite films are known to exhibit a slightly reduced
magnetization, which would lead to a smaller magnetostatic
energy).\cite{MPS+96,MPR+97,MGB+04} This value would imply that the
Fe$_3$O$_4$/\bto\ films would remain in-plane magnetized, but this
assumes a state of uniform strain; in real films, one may expect
larger interface strains that could give rise to larger
magnetoelastic anisotropies than those predicted in our simple model
calculations.

\subsection{CoFe$_2$O$_4$/\bto(100) epitaxial films}

We consider here the case when the CoFe$_2$O$_4$ film is fully
strained to the \bto\ substrate, a situation which is expected to
occur at thicknesses below the coherent growth thickness. Large
changes in magnetization as a function of temperature have been
reported for partially relaxed 50-150 nm CoFe$_2$O$_4$/\bto(100)
epitaxial films.\cite{CS06} Cobalt ferrite (CoFe$_2$O$_4$) is
characterized by very large magnetostriction coefficients and has
been a material of choice for strain-driven composite
multiferroics.\cite{ZWL+04} It is ferrimagnetic with a Curie
temperature of 790 K and is predicted to be nearly fully spin
polarized.\cite{PSSK92,AHY03}

\begin{figure}[t!bh]
\includegraphics[width=\columnwidth]{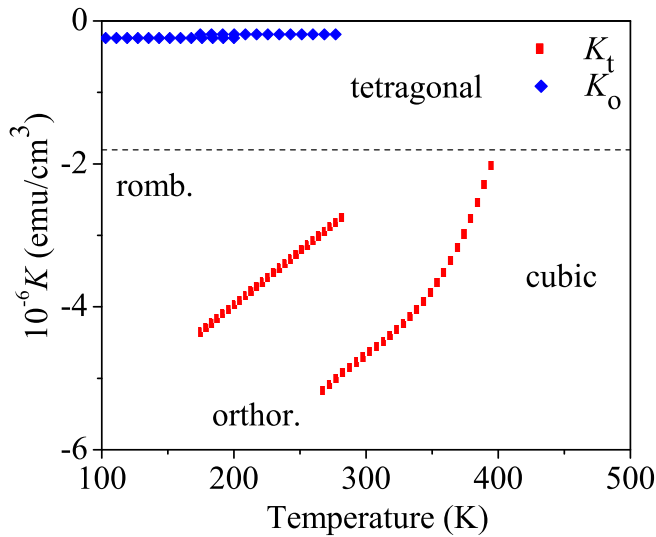}
\caption{Variation of the in-plane magnetoelastic anisotropy
constants of fully strained CoFe$_2$O$_4$/\bto(100). Diamonds
correspond to anisotropy in oblique domains, rectangles to
rectangular domains; the dashed line corresponds to the room
temperature magnetocrystalline anisotropy of bulk
Co$_{1.1}$Fe$_{1.9}$O$_4$ (which favors the $\langle 100 \rangle$
directions).} \label{fig:CoFe2O4}
\end{figure}

The calculated magnetoelastic anisotropies across the different
phases of \bto\ using the expressions derived above are shown in
Fig.~\ref{fig:CoFe2O4}, where we took into account the temperature
variation of the lattice parameter of
CoFe$_2$O$_4$.\cite{Taylor85,IKH+89} The out of plane anisotropies
for the fully strained film are very large, of the order of $9\times
10^{7}$ erg/cm$^3$, and favor in-plane magnetization, so we have
omitted these from the figure. In the cubic phase, the biaxial
strain contributes only to second order to the magnetic
anisotropy,\cite{VB01} and the in-plane magnetic anisotropy will be
dominated by the magnetocrystalline anisotropy, of the order of
$2\times 10^{6}$ erg/cm$^3$ (favoring the in-plane $\langle
100\rangle$ directions). In the tetragonal phase, the rectangular
domains induce a very large uniaxial anisotropy favoring the
direction of in-plane compression ($x$ axis in
Fig.~\ref{fig:strain}). In the orthorhombic phase, the distortion
induced by the rectangular domains is smaller, and the magnetic
anisotropy decreases slightly. The oblique distortions of the
orthorhombic and rhombohedral phases induce relatively small
magnetic anisotropies due to the much smaller value of
$\lambda_{111}$, and the magnetic anisotropy may be dominated by the
magnetocrystalline anisotropy, assuming that a value comparable to
the tabulated room temperature value still holds.

Magnetization versus temperature measurements of epitaxial
CoFe$_2$O$_4$/\bto(001) films have been presented by Chopdekar and
Suzuki\cite{CS06} for a 56 nm film under an applied magnetic field
$H = 50$ kOe, along the nominal [100] and [010] in-plane directions.
Even at this very large magnetic field, large jumps in the
magnetization are observed at the tetragonal to orthorhombic and
orthorhombic to rhombohedral phase transitions of \bto. As the
temperature is lowered from 350 K to 150 K, the magnetization for
$\mathbf{H} \parallel [010]$ direction increases slightly at 290 K,
and then decreases more drastically at 190 K; when $\mathbf{H}
\parallel [100]$ direction, the magnetization behaves in the
opposite fashion, starting and ending at values that are close to
those of the magnetization when the magnetic field is applied along
the [010] direction. These changes in the magnetization can be
understood by assuming predominantly square domains in the
tetragonal phase of \bto, which give similar anisotropies for both
the [100] and [010] directions; the presence of a small number of
rectangular domains, or a small imbalance in the two possible
rectangular domains, would give rise to a small uniaxial magnetic
anisotropy favoring, say, the [010] direction; these domains would
either remain rectangular in the orthorhombic phase or revert to
oblique domains. In the orthorhombic phase, the square domains
change to rectangular and the increase in magnetic energy along the
[100] direction (leading to a smaller magnetization) could result
from rectangular domains that are predominantly elongated along the
[100] direction (the easy axis direction is along the direction of
compression for CoFe$_2$O$_4$, which has $b_1>0$). The transition to
the rhombohedral phase would lead to similarly small magnetic
anisotropies for both [100] and [010] directions, leading to an
increase in magnetic energy for the [010] direction, and a decrease
along the [100] direction. Despite clear hysteresis in the
ferroelectric transitions of \bto, the observation of reproducible
paths in the magnetic behavior suggests that also the ferroelectric
domains of \bto\ also follow similar switching paths with thermal
cycling.

\subsection{Fe/\bto(100) epitaxial films}

\begin{figure}[t!]
\includegraphics[width=\columnwidth]{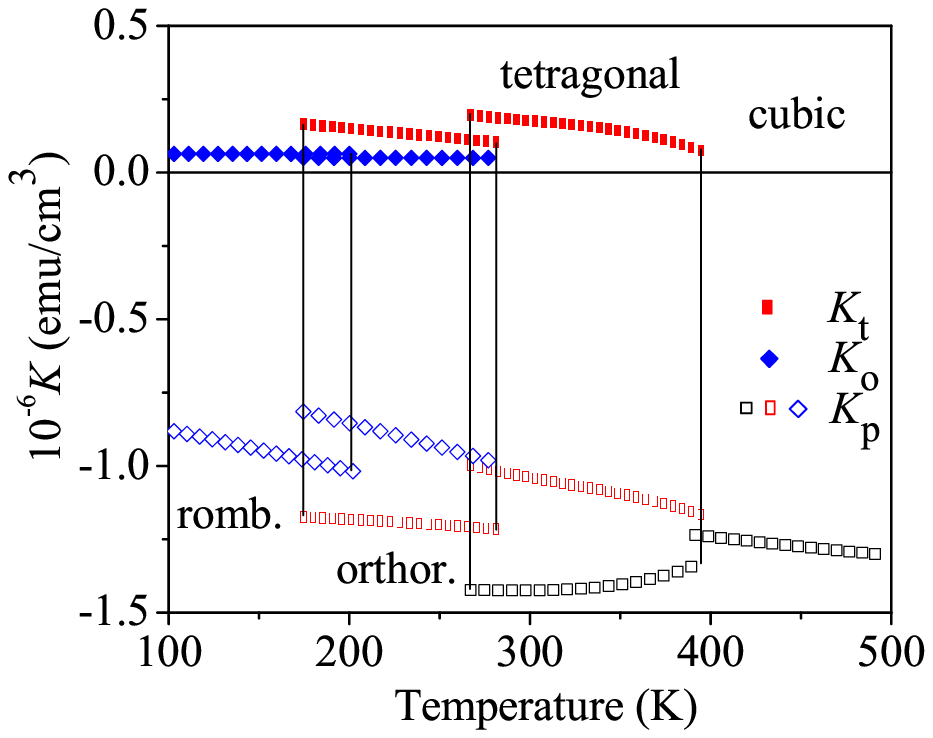}
\caption{Variation of the magnetoelastic anisotropy constants of
fully strained Fe/\bto(100). Square symbols correspond to anisotropy
in planar square domains, diamonds to oblique domains, and
rectangles to rectangular domains. Filled symbols are for in-plane
anisotropies and empty symbols are for the perpendicular anisotropy
constants.} \label{fig:Fe}
\end{figure}

As a final example, we consider the case of Fe, where the relatively
small lattice mismatch with \bto\ of --1.4\% should allow the growth
of fully strained epitaxial films. Further interest in this
particular system stems from recent theoretical predictions of
surface magnetoelectric effects induced by spin-polarized screening
at the Fe/\bto\ interface.\cite{DJT06} The calculated magnetoelastic
anisotropies for the different phases of \bto\ are shown in
Fig.~\ref{fig:Fe}. Due to the strong crystal field of metallic Fe,
the orbital moment is quenched, and the magnetic anisotropies are
relatively small in this system. It can be seen that for
Fe/\bto(100), the perpendicular magnetoelastic anisotropies favor
perpendicular magnetization, but given the very large magnetostatic
energy of Fe ($2\pi M_s^2 = 1.85 \times 10^7$ erg/cm$^3$), an
in-plane magnetized state is energetically more favorable. The
magnetoelastic in-plane anisotropies are also relatively small due
to the small strains involved, but are likely to strongly influence
the switching behavior of the Fe film.

Sahoo et al.\cite{SPD+07} have reported magnetic measurements on a
10 nm polycrystalline Fe film deposited on a \bto(100) substrate at
373 K, where the magnetization under a small applied magnetic field
(20-80 Oe) shows sudden jumps at the phase transitions of \bto.
Although the Fe film is likely to be fully relaxed, the relative
changes in magnetic anisotropy can still be compared qualitatively
with the results shown in Fig.~\ref{fig:Fe}. For the particular
field direction, which is applied in the plane of the film, the
magnetization is found to increase steeply with decreasing
temperature at both $\sim$278 K and $\sim$183 K, indicating a
decrease in the magnetic anisotropy along this direction. The
results shown in Fig.~\ref{fig:Fe} suggest that the magnetic field
is being applied along the $a$ direction of the rectangular domains
of the tetragonal phase, which corresponds to a hard axis direction
(in Fe this is the direction of compression, since $b_1< 0$) and
that the dominant planar domains in the orthorhombic phase of \bto\
are rectangular, compressed in this direction. This also agrees with
the measured temperature variation of the perpendicular
magnetization, where an increase at $\sim$278 K is followed by a
decrease at $\sim$183 K with decreasing temperature, in agreement
with the variation found for $K_\mathrm{p}$, shown in
Fig.~\ref{fig:Fe}.

\section{conclusions}

Explicit expressions for the magnetoelastic anisotropy constants
have been derived for arbitrary in-plane distortions of the unit
cell of cubic systems and applied to the particular case where those
distortions are induced by elastic coupling to \bto\ substrates.
\bto\ undergoes a series of phase transitions as a function of
temperature and has been employed to study the effect of strain on
the magnetic properties of epitaxial magnetic films and, in
particular, aimed at designing new magnetoelectric composite
systems. Since the elastic properties of ferroelectrics, and \bto\
in particular, can be modulated by electric fields via the
piezoelectric effect, modulation of magnetism via electric fields is
possible. We have considered three particular magnetic systems that
have been studied experimentally (Fe$_3$O$_4$, CoFe$_2$O$_4$, and
Fe), and explained the observed magnetic behavior in terms of the
changes in magnetic anisotropies arising from strain induced in the
different phases of \bto. This work provides a quantitative account
of the experimental observations and opens the way to a more
calibrated targeting of the properties of heterosystems for optimum
magnetoelectric coupling effects.

\begin{acknowledgments}
This work was supported by the NSF through MRSEC DMR 0520495
(CRISP).
\end{acknowledgments}


\end{document}